\documentstyle[aps,prl,epsf,twocolumn,floats]{revtex}

\begin{document}
\draft
\twocolumn[\csname @twocolumnfalse\endcsname
\widetext

\title{Quantum critical behavior in the Kondo problem with a pseudogap}
\author{Kevin Ingersent$^{(a)}$ and Qimiao Si$^{(b)}$}
\address{$^{(a)}$Department of Physics, University of Florida, Gainesville,
FL 32611-8440 \\
$^{(b)}$Department of Physics, Rice University, Houston, TX 77251-1892}
\date{19 October 1998}
\maketitle

\begin{abstract}

The variant of the single-impurity Kondo problem in which the
conduction-band density of states has a power-law pseudogap
at the Fermi energy is known to exhibit a zero-temperature phase
transition at a finite exchange coupling.
The critical properties of this transition are studied both for
$N=2$ and for $N\gg 1$, where $N$ is the spin degeneracy.
The critical exponents are consistent with a simple scaling
form for the free energy.
For any finite $N$, the temperature exponent of the local spin susceptibility
at the critical Kondo coupling varies continuously with the power of the
pseudogap.
This raises the possibility that a single-particle pseudogap is responsible
for the anomalous behavior of certain heavy-fermion metals close to a magnetic
quantum phase transition.

\end{abstract}
]

\narrowtext
A number of heavy-fermion metals exhibit non-Fermi-liquid behavior
which seems to arise from proximity to a zero-temperature
magnetic ordering transition \cite{ITP}.
These systems represent a puzzle in that temperature exponents describing
various transport and thermodynamic properties differ from the predictions
of the standard Ginzburg-Landau-type theory for spin fluctuations
\cite{Hertz,Millis}.
Recent neutron scattering experiments in CeCu$_{6-x}$Au$_x$
suggest that this discrepancy may arise from an unusual momentum
and energy dependence of the spin susceptibility \cite{Rosch,Schroder}.
In particular, an anomalous energy dependence extends over a wide range
of momenta, suggesting that the {\em local\/} spin susceptibility is
also anomalous \cite{Schroder}.
The microscopic origin of this local anomaly is unclear at present.

This paper explores the possibility that an anomalous local
susceptibility arises from a pseudogap in the single-particle spectrum.
(In quantum-critical systems which display superconductivity, such as
CePd$_2$Si$_2$ and CeIn$_3$ \cite{Mathur}, superconducting fluctuations may
generate a pseudogap \cite{pseudogap}.
Alternative mechanisms \cite{pseudogap} may be at work in other systems.)
To this end, we study a variant \cite{Withoff} of the SU($N$)-symmetric Kondo
model in which a single impurity moment (of spin degeneracy $N$) couples to a
conduction band described by the (oversimplified) density of states
\cite{energies}
\begin{eqnarray}
\rho(\epsilon) = \left\{
   \begin{array}{ll}
       \rho_0 |\epsilon|^r \quad & \mbox{for } |\epsilon| \le 1, \\[0.5ex]
       0 & \mbox{for } |\epsilon| > 1.
   \end{array} \right.
\label{dos}
\end{eqnarray}
In a metal ($r=0$), any antiferromagnetic Kondo coupling causes the
impurity moment to become quenched at sufficiently low temperatures
\cite{Hewson}.
With a power-law pseudogap ($r>0$), by contrast, quenching takes place
only if the Kondo coupling exceeds a nonzero critical value \cite{Withoff}.
This threshold is associated with a zero-temperature
phase transition, which is the focus of this work.

Our analysis is based on the observation that the critical behavior of the
power-law Kondo model reveals itself not in the conventional magnetic
susceptibility, but rather in the response to a local magnetic field coupled
solely to the impurity spin.
We propose a simple scaling form for the critical component of the free-energy,
which in turn leads to scaling relations among the critical exponents.
Within the large-$N$ approach \cite{Read,Cassanello}, we calculate the
critical exponents on the strong-coupling side, and then apply the scaling
ansatz to obtain the remaining exponents.
For $N=2$, all the exponents are computed directly using the numerical
renormalization-group (NRG) method \cite{Wilson,Chen,Gonzalez-Buxton},
and are found to obey the scaling relations to high accuracy.
Most importantly, the pseudogap fundamentally alters the local spin
susceptibility at the critical Kondo coupling, such that (for $N$ finite)
the critical exponent varies continuously with~$r$.

We begin with the Hamiltonian
\begin{eqnarray}
{\cal H}_{\text{K}} = \sum_{{\bf k},\sigma} \epsilon_{\bf k}
    c_{{\bf k}\sigma}^{\dagger} c_{{\bf k}\sigma}
  + {J_0 \over N} \sum_{\sigma,\sigma'} \vec{S} \cdot c_{0\sigma}^{\dagger}
    \vec{\tau}_{\sigma \sigma'} c_{0\sigma'},
\label{Kondo.hamiltonian}
\end{eqnarray}
where $J_0 > 0$ is the Kondo coupling, $c^{\dagger}_{0\sigma}$
creates a spin-$\sigma$ electron ($\sigma = 1, 2, \ldots, N$) at the impurity
site, and $\tau^i_{\sigma \sigma'}$ ($i=1, 2, \ldots, N^2-1$) is a generator of
SU($N$).
We employ a pseudofermion representation of the impurity spin operator,
$\vec{S}= N^{-1}\sum_{\sigma,\sigma'} f_{\sigma}^{\dagger}
\vec{\tau}_{\sigma \sigma'} f_{\sigma'}$, supplemented by a constraint
$\sum_{\sigma} f_{\sigma}^{\dagger} f_{\sigma} = Q$.
This paper treats systems possessing exact particle-hole symmetry,
for which $Q=N/2$;
the conduction-band dispersion $\epsilon_{\bf k}$ is assumed to give rise to
the density of states in Eq.~(\ref{dos}) with $0<r\le 1$.
More general cases will be described elsewhere.

To study the magnetic response, we add to Eq.~(\ref{Kondo.hamiltonian})
the magnetic field terms
\begin{equation}
{\cal H}_{\text{mag}} = \sum_{\sigma} \frac{m_{\sigma}}{2}
\left[ (H\!+\!h) f^{\dagger}_{\sigma} f_{\sigma'}
+ H \sum_{\bf k} c_{{\bf k}\sigma}^{\dagger}c_{{\bf k}\sigma} \right] .
\label{H_mag}
\end{equation}
Here $m_{\sigma} = \text{sgn}(2\sigma-N-1)$ and
$H$ represents the uniform magnetic field which couples with equal
$g$ factors to the impurity and the conduction-electron spins, while
$h$ is a local field coupling only to the impurity spin \cite{units}.

{\em Large-$N$ formulation:}
The large-$N$ limit of the model can be analyzed
within slave-boson mean-field theory \cite{Read,Withoff,Cassanello}.
In the path-integral representation of the partition function,
a Lagrange multiplier $\epsilon_f$ enforces the constraint on the impurity
occupancy, and Hubbard-Stratonovich field $\phi$ decouples the Kondo
interaction term.
The impurity part of the free energy, obtained by integrating out
the conduction and pseudofermion degrees of freedom, takes the form
\cite{units}
\begin{eqnarray}
F_{\text{imp}}&(&T,H,h) =
-\left[ {N \over 2 \pi}  \int_{-1}^1 d \epsilon
\; n_F(\epsilon\!+\!H/2) \, \delta_0 (\epsilon,h) \right] \nonumber\\
&& - \left[ \rule[-1ex]{0ex}{4ex} H\rightarrow -H, h \rightarrow -h \right]
+ \frac{N}{J_0} \phi^2 - Q \, \epsilon_f, \label{F_imp}
\end{eqnarray}
where $n_F$ is the Fermi function and $\delta_0$ is the conduction-band
phase shift:
\begin{equation}
\delta_0(\epsilon,h) = \tan^{-1}{\phi^2 \text{Im}\, G_0 (\epsilon+i0^+)
\over {\epsilon - \epsilon_f - h/2 - \phi^2 \text{Re}\, G_0 (\epsilon+i0^+)}} .
\label{delta_0}
\end{equation}
The noninteracting local conduction electron Green's function satisfies
$\text{Im}\,G_0(\epsilon+i0^+) = -\pi\rho(\epsilon)$,
and for $|\epsilon|\ll 1$ \cite{Withoff,Cassanello},
\[
\text{Re}\,G_0(\epsilon\!+\!i0^+) \approx \left\{
\begin{array}{ll}
-\pi\rho(\epsilon) \tan\!\left({\displaystyle\frac{r\pi}{2}}\right)
\text{sgn} \, \epsilon , \; & 0 < r < 1, \\[1.5ex]
-2\rho(\epsilon) \ln|\epsilon^{-1}|, & r = 1.
\end{array} \right.
\]

The mean-field solution is obtained by minimizing $F_{\text{imp}}$ with
respect to $\epsilon_f$ and $\phi^2$ at zero magnetic field.
Particle-hole symmetry ensures that $\epsilon_f=0$,
while $\phi^2$ is determined by the saddle-point equation
\begin{equation}
{1 \over J_0} = \int_{-1}^{1} {{d \epsilon}\over \pi}
{{ n_F (\epsilon) \, (\epsilon - \epsilon_f) \, \text{Im}\, G_0 (\epsilon)}
\over [\epsilon - \epsilon_f - \phi^2 \text{Re}\, G_0(\epsilon)]^2
+ [ \phi^2 \text{Im}\, G_0 (\epsilon )]^2},
\label{saddle.point.eq}
\end{equation}
which is exact to $O(1/N)$.

The critical coupling $J_c$, extracted
by the condition that $\phi$ vanishes as $J_0\rightarrow J_c^+$
from Eq.~(\ref{saddle.point.eq}) with $T=H=h=0$,
satisfies $\rho_0 J_c=r$ \cite{Withoff}.
Just above the critical coupling, we find that \cite{r=1/2}
\begin{equation}
\phi^2 \propto (J_0 - J_c)^{1/\bar{r}-1}, \quad
\bar{r}=\min(r,{\textstyle\frac{1}{2}}).
\label{phi(J)}
\end{equation}

{\em Local vs impurity susceptibility:}
An important clue to the nature of the phase transition for $r>0$ comes from
the contrast between the zero-temperature limits of the impurity susceptibility
$\chi_{\text{imp}} = -\partial^2 F_{\text{imp}}/\partial H^2|_{H=h=0}$
and the local susceptibility
$\chi_{\text{loc}} = -\partial^2 F_{\text{imp}}/\partial h^2|_{H=h=0}$.
NRG results \cite{Chen,Gonzalez-Buxton} for degeneracy $N=2$ indicate
that whereas $\lim_{T\rightarrow 0}T\chi_{\text{imp}}$ is nonzero for all
values of $J_0$, $\lim_{T\rightarrow 0}T\chi_{\text{loc}}$ goes continuously
to zero as the critical coupling is approached from below and
$\lim_{T\rightarrow 0}T\chi_{\text{loc}}=0$ for all $J_0>J_c$.
We now show that in most cases, the same distinction can be made in the
large-$N$ limit.

Starting from Eq.~(\ref{F_imp}), the impurity susceptibility can be
written
\begin{equation}
\chi_{\text{imp}} = \frac{N}{4\pi} \int_{-1}^1 d \epsilon \;
\frac{d n_F(\epsilon)}{d\epsilon} \,
\frac{\partial\delta_0(\epsilon,0)}{\partial\epsilon} .
\end{equation}
 From Eq.~(\ref{delta_0}),
$\delta_0(\epsilon\!\rightarrow\!0,0) \approx
(1\!+\!r\,\text{sgn}\,\epsilon) \pi / 2$ \cite{jump},
so
\begin{equation}
\lim_{T \rightarrow 0} T \chi_{\text{imp}} = N r / 16.
\label{chi_imp}
\end{equation}
This large-$N$ result reduces in the case of $N=2$ to the value
$r/8$ obtained within the NRG approach \cite{coincidence}.

The local spin susceptibility can be evaluated directly by differentiating
Eq.~(\ref{F_imp}).
Two very different types of behavior are predicted, depending on the
value of $r$.
For $0 < r < \frac{1}{2}$, we find that
\begin{equation}
\chi_{\text{loc}}(J_0>J_c, T=0) \propto (J_0 - J_c)^{-1/r}.
                                                \label{chi_loc}
\end{equation}
This is consistent with the result obtained for $N=2$, namely
$\lim_{T\rightarrow 0} T\chi_{\text{loc}}=0$.
Note that unlike $\chi_{\text{imp}}$, which is independent of $J_0$
throughout the strong-coupling regime, $\chi_{\text{loc}}(T=0,J_0)$
diverges as $J_0\rightarrow J_c^+$.

For $\frac{1}{2}\le r\le 1$, the large-$N$ result is $\chi_{\text{loc}}=\infty$.
This unusual behavior has a counterpart in the case $N=2$, where
$J_c=\infty$, i.e., the strong-coupling regime is completely inaccessible
\cite{Gonzalez-Buxton}.
Away from strict particle-hole symmetry, a critical $\chi_{\text{loc}}$
is recovered for all $r>0$;
however, the treatment of particle-hole asymmetry lies beyond the scope of the
present paper, so we focus henceforth on the range $0 < r < \frac{1}{2}$.

{\em General scaling analysis:}
In the standard ($r=0$) Kondo problem, the zero-temperature phase transition
between weak and strong coupling has the Kosterlitz-Thouless form.
Based on the Coulomb-gas picture of the partition function\cite{Hewson}, one
expects that the Kosterlitz-Thouless nature of the transition is special
to $r=0$. For $r>0$, by contrast, it is natural to assume that the transition
is continuous (second-order).

At a continuous transition, the singular component of the free energy $F_s$
should exhibit scaling.
The deviation of the Kondo exchange from its critical value, $J_0-J_c$, is
clearly a relevant variable. The different behaviors of $\chi_{\text{imp}}$ and
$\chi_{\text{loc}}$ reported above suggest that the local magnetic field~$h$
(rather than the uniform magnetic field~$H$) should act as a second scaling
variable.
$F_s$ is therefore expected to have the form
\begin{eqnarray}
F_{s} = T f\left(|J_0\!-\!J_c|/T^a, \, |h|/T^b\right).
\label{F_s}
\end{eqnarray}

Alternatively, one can characterize the critical behavior by a set of exponents
$\beta$, $\gamma$, $\delta$, and $x$, defined as follows:
\begin{eqnarray}
M_{\text{loc}}(J_0 < J_c,T=0,h=0)
&\propto& (J_c-J_0)^{\beta}, \nonumber\\
\chi_{\text{loc}}(J_0 > J_c,T=0) &\propto& (J_0-J_c)^{-\gamma},
\nonumber\\[-1.5ex]
\label{exponents} \\[-1.5ex]
M_{\text{loc}}(J_0=J_c,T=0) &\propto& | h |^{1/\delta}, \nonumber\\
\chi_{\text{loc}}(J_0=J_c) &\propto& T^{-x}. \nonumber
\end{eqnarray}
The local magnetization
$M_{\text{loc}}(h\!=\!0) = - \partial F_s/\partial h |_{h=0}$ acts as the order
parameter for the transition and
$\chi_{\text{loc}} = - \partial^2 F_s/\partial h^2 |_{h=0}$
is the order-parameter susceptibility.
The critical exponents can readily be written in terms of $a$ and $b$
introduced in Eq.~(\ref{F_s}):
$\beta = (1-b)/a$,
$\gamma = (2b-1)/a$,
$\delta = b/(1-b)$,
and $x = 2 b - 1$.
These expressions lead to a pair of scaling relations
among the critical exponents, e.g.,
\begin{mathletters}
\begin{eqnarray}
\delta &=& (1 + x) / (1 - x), \label{scaling_dx} \\
\beta &=& \gamma (1 - x) / (2 x). \label{scaling_bgx}
\end{eqnarray}%
\label{scaling}%
\end{mathletters}
\hspace{\parindent}
{\em Critical behavior in the large-$N$ limit:}
Equation~(\ref{chi_loc}) implies that $\gamma=1/r$ for $0 < r < \frac{1}{2}$.
Since the large-$N$ mean-field solution is valid only for $J_0>J_c$,
it is not possible to determine the exponents $x$, $\delta$, and $\beta$
directly.
These exponents may be inferred, however, from the general scaling analysis
outlined above.
Combining Eqs.~(\ref{F_imp})--(\ref{phi(J)}), we find that \cite{r=1/2}
\begin{equation}
F_s(J_0>J_c,T=H=h=0) = -N A (J_0-J_c)^{1/\bar{r}},
\label{F_s1}
\end{equation}
where $A$ is an $r$-dependent coefficient and $\bar{r}=\min(r,\frac{1}{2})$
as before.
If we assume the validity of Eq.~(\ref{F_s}), we can deduce that $a=r$ and
$b=1$ for $0 < r < \frac{1}{2}$, and hence that the full set of exponents is
\begin{equation}
\gamma = 1/r, \quad \beta = 1/\delta = 0, \quad x = 1.
\label{large_N_exps}
\end{equation}
Only $\gamma$ shows any $r$-dependence at the mean-field level.

We note that $-F_s(T=0)/N=T_K$, the Kondo scale (the
inverse correlation length in the temporal direction).
The vanishing of $T_K$ in power-law fashion as $J_0$ approaches $J_c$
[see Eq.~(\ref{F_s1})] is precisely the behavior expected at a second-order
phase transition, and confirms that the critical point for $r>0$ is not of the
Kosterlitz-Thouless type.

{\em Critical behavior for $N=2$:}
We treat the case $N=2$ using a generalization of Wilson's NRG method
to describe a nonconstant density of states \cite{Gonzalez-Buxton}.
The local magnetization and local susceptibility can be computed as
$M_{\text{loc}}=\langle S_z\rangle$ and
$\chi_{\text{loc}}=\lim_{h\rightarrow 0}\langle S_z\rangle / h$,
where $\langle S_z\rangle$ is the expectation value of the $z$ component
of the impurity spin.
We have performed calculations for several different values of $r$
between $0.1$ and $0.45$.
(As mentioned above, the strong-coupling regime disappears for
$r\ge\frac{1}{2}$.)
In each case $M_{\text{loc}}$ and $\chi_{\text{loc}}$
behave as described in Eqs.~(\ref{exponents}),
establishing the continuous nature of the phase transition.

\begin{table}[t]
\begin{tabular}{llllll}
\multicolumn{1}{c}{$r$} & \multicolumn{1}{c}{$\Lambda$} &
\multicolumn{1}{c}{$\beta$} & \multicolumn{1}{c}{$\gamma$} &
\multicolumn{1}{c}{$1/\delta$} & \multicolumn{1}{c}{$1-x$} \\[0.5ex]
\hline\\[-1ex]
0.1  & 9 &          &         & 0.005651(3) & 0.0112(1)  \\
0.15 & 9 & 0.095(5) & 7.14(9) & 0.01367(3)  & 0.0270(1)  \\
0.2  & 9 & 0.153(2) & 5.81(5) & 0.02636(4)  & 0.0515(1)  \\
0.3  & 9 & 0.35(1)  & 4.39(2) & 0.07405(5)  & 0.1378(5)  \\
     & 3 &          &         &             & 0.13715(5) \\
0.4  & 9 & 0.90(1)  & 3.90(2) & 0.1850(2)   & 0.3124(9)  \\
     & 3 &          & 3.98(5) &             & 0.3113(2)  \\
0.45 & 9 & 1.91(2)  &         & 0.3135(3)   & 0.4771(9)
\end{tabular}
\vspace{2ex}
\caption{
Critical exponents for the case $N=2$, obtained from NRG calculations.
Parentheses indicate the estimated random error in the last digit.
Here $\Lambda$ parameterizes the discretization of the conduction band.
}
\label{tab:exps}
\end{table}

Table~\ref{tab:exps} lists the exponents extracted from the NRG runs
along with their estimated random errors.
Data at the critical point exhibit asymptotic power-laws over at least five
decades of $T$ or $h$, resulting in rather precise determinations of $x$ or
$\delta$, respectively.
The determination of the other two exponents is hindered by numerical
rounding errors which eventually cut off the power-law variation of
$M_{\text{loc}}$ and $\chi_{\text{loc}}$ on approach to the critical
coupling.
Some of the $\beta$ and $\gamma$ values in Table~\ref{tab:exps} have been
computed from data taken over only one decade of $|J_0 - J_c|$.
This problem becomes even more pronounced as $r$ approaches $\frac{1}{2}$,
and we have found it impossible to make a reliable estimate of $\gamma$ for
$r>0.4$.
Very small values of $r$ also prove to be inaccessible due to the
buildup of rounding error during the approach to the limit $T=0$.

Most runs were performed for an NRG discretization parameter $\Lambda=9$,
retaining all many-body eigenstates up to an energy $50T$ above the
ground state.
In an attempt to estimate the systematic discretization errors, selected
runs were performed using $\Lambda=3$, a value which lies closer to the
continuum limit ($\Lambda=1$) but which requires much more computer time.
It can be seen from Table~\ref{tab:exps} that those exponents computed for
both $\Lambda=3$ and $\Lambda=9$ narrowly fail to agree within their estimated
random errors.
We therefore believe that the $\Lambda=9$ exponents provide a
fair approximation to the continuum values.

All four exponents listed in Table~\ref{tab:exps} show significant
$r$-dependence.
In all cases, they satisfy the general scaling relations very well.
Equation~(\ref{scaling_dx}) relating $\delta$ and $x$ is
obeyed to better than 0.5\% in absolute terms, and to well within the
estimated errors in the exponents.
Equation~(\ref{scaling_bgx}) is obeyed to within 5\%, a level
consistent with the greater uncertainties in the values of $\beta$
and $\gamma$.
This agreement supports the validity of the scaling hypothesis presented in
Eq.~(\ref{F_s}).

\begin{figure}
\centering
\vbox{\epsfxsize=65mm \epsfbox{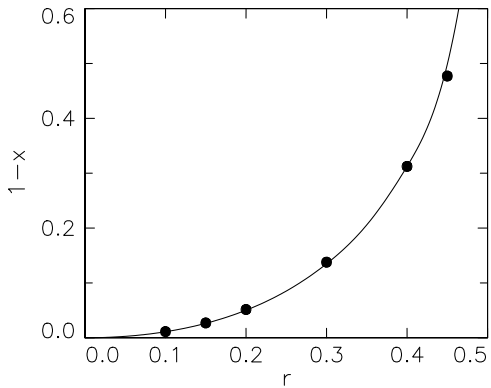}}
\vspace{1ex}
\caption{
Exponent $1-x$ plotted vs $r$ for $N=2$.
The symbols plot the $\Lambda=9$ data from Table~\protect\ref{tab:exps}.
(Estimated errors are smaller than the symbols.)
The solid line shows the perturbative prediction of
Eq.~\protect(\ref{x:small_r}), in the form of a spline fit through numerical
values for $(\rho_0 J_c)^2$, calculated for $\Lambda=3$ and extrapolated to
the continuum limit $\Lambda=1$ (see Ref.~\protect\onlinecite{Gonzalez-Buxton}).
}
\label{fig:x}
\end{figure}

{\em Small-$r$ expansion:}
There is an alternative method for determining the exponent $x$ associated
with the local spin susceptibility at the critical Kondo-coupling.
The local spin-spin correlation function can be calculated perturbatively in
terms of $\rho_0J_c\approx r$, in analogy to the standard $\epsilon-$expansion.
Exponentiating the leading logarithmic term in the perturbation series
leads to
\begin{eqnarray}
x = 1 - \frac{2}{N}(\rho_0 J_c)^2, \quad \rho_0 J_c \ll 1.
\label{x:small_r}
\end{eqnarray}
Figure~\ref{fig:x} compares this result with the NRG data for $N=2$.
The agreement is remarkably good, even when $r$ (and hence $\rho_0 J_c$)
is not small.

For large but finite $N$, inserting the slave-boson mean-field value
$\rho_0 J_c=r$ into Eq.~(\ref{x:small_r}) yields $x = 1 - 2 r^2/N + O(1/N)^2$.
For $N=\infty$, this expression reproduces that obtained by combining
slave-boson mean-field theory with the scaling ansatz
[see Eq.~(\ref{large_N_exps})].
The agreement between the two results provides additional support for
the scaling hypothesis.

Finally, we note that the many-body spectrum at the critical point displays
a non-Fermi liquid form.
This is to be expected: since $J_c \neq 0$, interactions do not renormalize
to zero.
We have also studied the particle-hole-asymmetric problem, and
again found the exponent~$x$ to be $r$-dependent.
Details will be given elsewhere.

In summary, we have analyzed the critical properties of the Kondo problem
with a pseudogap in the conduction-electron density of states.
We find that the temperature exponent of the local spin susceptibility at
the critical point depends continuously on the power of the pseudogap.
(For large $N$, such a dependence appears only at order $1/N$.)
The implications of our results for lattice heavy fermion systems can be
seen within the dynamical mean-field approach \cite{RMP}, which analyzes
a lattice problem in terms of an impurity coupled self-consistently to a
fermionic bath. The development of a pseudogap in the lattice system will
lead to a pseudogap in the corresponding effective impurity problem.
Should the self-consistency place the Kondo coupling near its critical value,
our results then imply that the local dynamical spin susceptibility
will have an anomalous exponent.
Tunneling and photoemission experiments in the relevant temperature
regime would provide a direct test for the existence of a pseudogap.

We would like to thank C.\ R.\ Cassanello, A.\ Chubukov, P.\ Coleman,
and E.\ Fradkin for useful discussions.
This work has been supported in part by NSF Grant No.\ DMR-9316587 (K.I.),
and by NSF Grant No.\ DMR-9712626 and an A.\ P.\ Sloan Fellowship (Q.S.).


\begin{references}
\bibitem{energies}
    All energies are expressed in units of the half-bandwidth and measured
    from the Fermi energy, $\epsilon = 0$. The condition
    $\int_{-1}^{+1}\!d\epsilon\,\rho(\epsilon)=1$ implies that
    $\rho_0=2/(1+r)$.
\bibitem{units}
    We work with units in which $k_B=1$ and $g\mu_B=1$.
\bibitem{jump}
    Alternative expressions for $\delta_0(\epsilon,0)$ arise from different
    choices as to which one of the strong-coupling levels in each of the $N$
    bands evolves from an impurity level, and hence is not described by a
    phase shift.
    The choice $\delta_0(\epsilon,0)=(r\!-\!1)(\pi/2)\,\text{sgn}\,\epsilon$
    makes explicit the particle-hole symmetry of the problem, but necessitates
    the addition to Eq.~(\ref{F_imp}) of a term representing the free
    energy of $N$ levels located at $\epsilon=0$.
    All expressions for $\delta_0(\epsilon,0)$ yield the same impurity
    contribution to any physical property.
\end{references}
\end{document}